# Phonon dynamics in $Sr_{0.6}K_{0.4}Fe_2As_2$ and $Ca_{0.6}Na_{0.4}Fe_2As_2$

R. Mittal[1,2], Y. Su[1], S. Rols[3], M. Tegel[4], S. L. Chaplot[2], H. Schober[3], T. Chatterji[5], D. Johrendt[4] and Th. Brueckel[1,6]

[1]*Juelich Centre for Neutron Science, IFF, Forschungszentrum Juelich, Outstation at FRM II, Lichtenbergstr. 1, D-85747 Garching, Germany*
[2]*Solid State Physics Division, Bhabha Atomic Research Centre, Trombay, Mumbai 400 085, India*
[3]*Institut Laue-Langevin, BP 156, 38042 Grenoble Cedex 9, France*
[4]*Department Chemie und Biochemie, Ludwig-Maximilians-Universitaet Muenchen, Butenandtstrasse 5-13 (Haus D), D-81377 Muenchen, Germany*
[5]*Juelich Centre for Neutron Science, Forschungszentrum Juelich, Outstation at Institut Laue-Langevin, BP 156, 38042 Grenoble Cedex 9, France*
[6]*Institut fuer Festkoerperforschung, Forschungszentrum Juelich, D-52425 Juelich, Germany*

We report inelastic neutron scattering measurements of the phonon density-of-states in superconducting $Sr_{0.6}K_{0.4}Fe_2As_2$ ($T_c$=30 K) and $Ca_{0.6}Na_{0.4}Fe_2As_2$ ($T_c$=18 K). Compared with the parent compound $BaFe_2As_2$ doping affects mainly the lower and intermediate frequency part of the vibrations. Mass effects and lattice contraction cannot explain these changes. A lattice dynamical model has been used to identify the character of the various phonon bands. Softening of phonon modes below 10 meV has been observed in both samples on cooling from 300 K to 140 K. In the Ca doped compound the softening amounts to about 1 meV while for the Sr doped compound the softening is about 0.5 meV. There is no appreciable change in the phonon density of states when crossing $T_c$.

The recent discovery of superconductivity in Fe-As layered structure compounds has attracted a huge attention [1-21] in the scientific community. In general terms these systems can be classified in two different families. One of these is derived from the parent compounds RFeAsO (R = La, Sm, Ce, Pr, Nd, Gd). Depending on the lanthanide ion, $T_c$ can be as high as 55 K on doping with F at the O site [2]. The highest $T_c$ of 56.5 K in iron-based superconductors so far was achieved in $Gd_{1-x}Th_xOFeAs$ without F doping [3]. The second type of family is derived from parent compounds of the composition $MFe_2As_2$ (where M=Ba, Sr, Eu, Ca). $T_c$ values as high as 38 K are found on partial substitution of Ba by K atoms [4]. The difference between the two families resides principally in the fact that in $MFe_2As_2$ the Fe-As layers are separated by M atoms while in RFeAsO the separation is achieved via R-O layers.

Extensive efforts have been undertaken to raise $T_c$ and to understand the mechanism of superconductivity in these compounds. Electronic structure calculations [9] show that in FeAs compounds the electronic bands around the Fermi level are formed mainly by Fe and As states, while the bands of La-O or M atoms are far from the Fermi level. It is only natural to believe that superconductivity in these compounds is due to the structural and electronic properties of the Fe-As layers. Recent inelastic neutron scattering measurements carried out on $BaFe_2As_2$ [10] and $Ba_{0.6}K_{0.4}Fe_2As_2$ [11] show the evidence of presence of magnetic excitations. Therefore, spin fluctuations may play an important role for the mechanism of superconductivity. However, it is equally demonstrated that phonons couple selectively to the spin system [12]. Therefore, and despite the fact that simple electron-phonon coupling mechanisms seem to be rather unlikely [13], it is extremely important to investigate the phonon spectrum experimentally, in order to clarify the role of electron-phonon coupling in both the spin-density wave ordering and the superconducting phase transition. Electron-phonon coupling may be reflected via the changes of phonon life time and phonon energies. The phonon frequency shift, especially when it occurs abruptly along a dispersion curve is perhaps the most readily accessible and well-known effect (Kohn anomaly) of electron-phonon coupling on the phonon system.

The As ion has a greater electronic polarizability in comparison to divalent oxygen. Thus these compounds are expected to be more compressible in comparison to copper based systems. The application of pressure has been shown to suppress the structural and magnetic transitions in these compounds [5-8]. The phase transition to a collapsed tetragonal phase and superconductivity seem to be related in $MFe_2As_2$. Application of pressure causes a steep increase in the $T_c$ of F-doped LaOFeAs from 27 K to 43 K at 4 GPa [5]. In $CaFe_2As_2$ pressure-induced superconductivity [6,7] was found at 0.35 GPa, while superconductivity for $BaFe_2As_2$ and $SrFe_2As_2$ happens [8] at significantly higher pressures of about 3.8 GPa and 3.2 GPa respectively. Polarized Raman spectra of non-superconducting $CaFe_2As_2$, $SrFe_2As_2$ and superconducting $Sr_{0.6}K_{0.4}Fe_2As_2$ have been communicated [15]. The lattice dynamics of $LaFeAsO_{1-x}F_x$ and $PrFeAsO_{1-y}$ has also been investigated via inelastic X-Ray scattering and first-principles calculations [16]. The experimental phonon data on LaFeAsO are not fully explained by ab-initio calculations [17].

We are exploring these compounds using the techniques of inelastic neutron scattering and lattice dynamical model calculations. Electron-phonon coupling should be detectable in the changes upon doping as well as in the temperature evolution of the spectra. The results presented here for superconducting $Sr_{0.6}K_{0.4}Fe_2As_2$ and $Ca_{0.6}Na_{0.4}Fe_2As_2$ build on our earlier investigation of the parent compound $BaFe_2As_2$ [19], which we will use as a reference system.

The polycrystalline samples of $Sr_{0.6}K_{0.4}Fe_2As_2$ ($T_c$=30 K) and $Ca_{0.6}Na_{0.4}Fe_2As_2$ ($T_c$=18 K) were prepared by heating stoichiometric mixtures of the corresponding purified elements. All samples were prepared in batches of 3 to 4.5 grams, heated and annealed several times in sealed niobium tubes under an atmosphere of purified argon. After each annealing step, the mixtures were homogenized in an agate mortar and pressed into pellets before the last annealing step. The samples were heated to 1073 – 1173 K ($Sr_{0.6}K_{0.4}Fe_2As_2$) and 973 – 1073 K ($Ca_{0.6}Na_{0.4}Fe_2As_2$) in the different annealing steps and kept at these temperatures for 30 to 48 h. In the first step, the mixtures were heated very slowly in the temperature range from 573 to 873 K and kept at this temperature for 12 h in order to prevent undesirable reactions. The inelastic neutron scattering experiments were performed using the IN4 and IN6 time of flight spectrometers at the Institut Laue Langevin (ILL), France. Measurements were made on about 8 grams of polycrystalline samples. The incoherent approximation [22] has been used for extracting neutron weighted phonon density of states from the measured scattering function $S(Q,E)$.

Using IN4 we have carried out measurements at 2.5 K and 50 K (Fig. 1). An incident neutron wavelength of 1.18 Å is used to cover the full phonon spectra in the energy range up to 45 meV. The experimental S(Q,E) measured for both compounds are shown in Fig. 1. The inelastic neutron scattering measurements carried out on powder samples of $BaFe_2As_2$ and $Ba_{0.6}K_{0.4}Fe_2As_2$ indicate neutron scattering evidence of a resonant spin excitation. In the dynamical range chosen for our phonon scattering measurements no detectable signs of such resonant spin excitations are observed. The phonon spectra obtained from the S(Q,E) under the incoherent approximation are shown in Fig. 2. They show only moderate changes in the density of states when going from above (50 K) to below the superconducting transition temperature (2.5 K). The formation of Cooper pairs thus seems to have only a very minor influence on the overall vibration spectrum.

The lattice dynamical calculations applied to analyze the experimental spectra follow closely those for $BaFe_2As_2$ reported previously [19] by us. The parameters of the interatomic potential are the effective charge and the radius of the atoms. The radii parameter corresponding to Sr/K and Ca/Na are chosen such that the potentials satisfy the conditions of static and dynamic equilibrium [23,24]. The radii parameters for Sr/K and Ca/Na site atoms are R(Sr/K) = 2.11 Å and R(Ca/Na) = 2.08 Å, respectively. All other parameters of the potential are the same as reported for $BaFe_2As_2$. The calculations have been carried out using the current version of the software DISPR [25] developed at Trombay.

To simulate the non-stochiometry arising from the doping the calculations are carried out for a $2 \times 2 \times 2$ super cell, where 60% of the M(Sr, Ca) atoms are randomly replaced by either K or Na atoms in the SrK and CaNa compounds, respectively. The calculated phonon spectra are also shown in Fig. 2. Calculations compare reasonably well with the experimental data from IN4 except for one peak at 21 meV and 17 meV in the Sr and Ca compounds respectively. This reflects the difficulties already encountered with a peak in this energy range in $BaFe_2As_2$. With respect to $BaFe_2As_2$ we find that the peak under question is much weaker in the Sr compound, while it is much more intense in the Ca compound. The partial densities of states (Fig. 3) give the dynamical contributions to frequency distribution arising from the various species of atoms. The Sr/K and Ca/Na atoms mainly contribute in the 0–25 meV range, while the As and Fe atoms contribute in the whole 0–40 meV range. Above 30 meV the contributions are mainly due to Fe-As stretching modes. Below 15 meV the contributions arise mainly from the Ca/Na and Sr/K atoms. The first low energy peak in the partial density of states (Fig. 3) of Ca/Na and Sr/K is at about 8 meV.

The superconducting properties of $MgB_2$ with a $T_c$ of 38 K, i.e. similar to that of the pnictide superconductors, have been explained in the framework of standard electron-phonon coupling. However, the phonons mediating the pairing are very hard. Their involvement in superconductivity is reflected in the strong change they experience upon doping [27]. In the present case the high-frequency band at 34 meV reacts relatively weakly to doping. In addition these small changes can be related to minor variations in the bond lengths. Therefore, and despite the fact that a BCS like gap has been found [18] in the superconductor $SmFeAsO_{0.85}F_{0.15}$ by Andreev spectroscopy, the high $T_c$ of the pnictides seems incompatible with standard electron-phonon coupling alone.

We have complemented the phonon density of states measurements at low temperatures with measurements of higher resolution in the low-frequency range (0 – 20 meV) using the IN6 spectrometer at a wavelength of 5.1 Å (Fig. 4). Due to their nature (up-scattering) these measurements are limited to higher temperatures. The phonon density of states for the Ba, CaNa and SrK compounds (Fig. 4) measured at 300 K using IN6 show pronounced differences in the lower half of the spectral range. It is not evident how to attribute these changes to a simple mass renormalization of the modes involving Ba (m= 137.34 amu), Sr (m= 87.62 amu) and Ca (m=40.08 amu). Qualitatively our data show in particular that the peak found at about 21 meV in the Ba and Sr compounds has been shifted to a lower energy of about 17.5 meV in the Ca compound. If we compare the unit cell dimensions of the CaNa, Ba and SrK compounds then we notice that all three compounds feature nearly the same value for the lattice parameter *a*, while the lattice parameter *c* in the CaNa compound is about 10 % shorter in comparison of that of the Ba and SrK compounds. The contraction of the unit cell should normally result in shifting the modes of the AsFe-planes, if at all, towards higher energies. The fact that the contrary is observed indicates that the bonding is quite different in the CaNa compound in comparison to Ba and SrK. The buffer layers are thus not just charge reservoirs.

Apart from the doping dependence the temperature dependence of the phonon spectra may give valuable insight into the dynamics of a superconductor. Usually phonon modes are found to shift towards higher energies with a decrease of the unit cell volume with decreasing temperature. This is actually what we observe for the high-frequency band centered on 34 meV when going from 300 to 140 K. The band around 25 meV shows a more complex behavior. It equally displaces the center-of-gravity to higher frequencies. However, it simultaneously narrows in width. Contrary to the high-frequency bands the low-energy phonon modes up to about 10 meV soften (Fig. 4) for both the CaNa and SrK compounds as we decrease the temperature from 300 to 140 K. In the CaNa doped compound the softening amounts to about 1 meV while for the SrK doped compound the softening is about 0.5 meV. A similar softening of low-frequency modes has e.g. been observed in superconducting $LuNi_2B_2C$ [26].

In general it is not easy to make a connection between temperature induced softening and electron-phonon coupling. In the present case the behavior of the system under pressure may give an indication. Superconductivity in the parent $MFe_2As_2$ compounds emerges when there is suppression of the structural phase transition and its concomitant magnetic ordering. The large volume collapse of $CaFe_2As_2$ [7] at a temperature corresponding to its pressure induced superconductivity indicates that the phase transition is electronic in origin. This strong correlation of electronic properties and structure would be compatible with a softening of low-energy phonon modes in CaNa and SrK compounds due to electron-phonon coupling (lowering of temperature corresponding structurally to an increase in pressure). The higher softening in the CaNa compound may indicate that electron-phonon coupling is stronger than for SrK. Since the phonon softening is observed only in the normal state of both superconducting samples, it is unlikely directly associated to superconductivity, and this despite the observed isotope effect in pinictide superconductors suggests [14]. The tetragonal to orthorhombic phase transition is suppressed in the superconducting compounds, the structural phase transition thus also appears not relevant to the observed phonon softening.

In summary, applying a combination of experimental phonon studies and lattice dynamical calculations for the superconducting compounds $Sr_{0.6}K_{0.4}Fe_2As_2$ and $Ca_{0.6}Na_{0.4}Fe_2As_2$ leads to a good understanding of the phonon spectra. Doping affects mainly the lower and intermediate frequency part of the vibrations. In particular the region around 20 meV that had already retained our attention in the parent compound $BaFe_2As_2$ shows a very strong renormalization. Mass effects and lattice contraction cannot explain these changes. Therefore, the type of buffer ion influences the bonding in the Fe-As layers. The buffers thus cannot be considered a mere charge reservoir. The high-frequency band reacts moderately to doping. In both compounds the low energy phonon modes soften with temperature. This softening, which is stronger in the CaNa compound, might be due to electron-phonon coupling effects. No anomalous effects are observed in the phonon spectra when passing the superconducting transition temperature. All this indicates that while electron-phonon coupling is present it cannot be solely responsible for the electron pairing.

Fig. 1. (Color online) The experimental S(Q,E) plots for $Ca_{0.6}Na_{0.4}Fe_2As_2$ and $Sr_{0.6}K_{0.4}Fe_2As_2$ at 2.5 K and 50 K measured using the IN4 spectrometer at the ILL with an incident neutron wavelength of 1.18 Å.

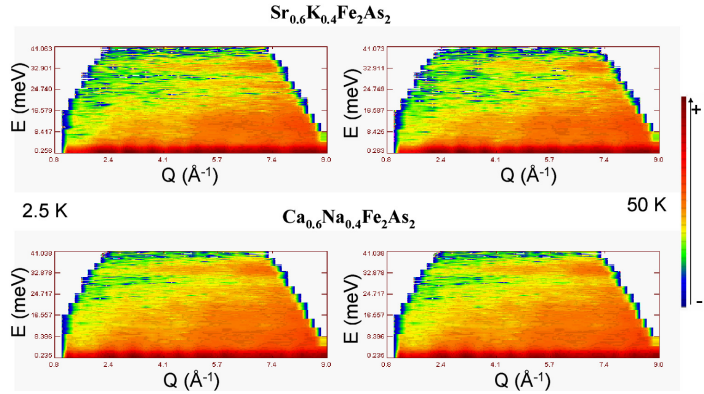

Fig. 2. (Color online) Comparison between the calculated and experimental phonon spectra of $Sr_{0.6}K_{0.4}Fe_2As_2$ and $Ca_{0.6}Na_{0.4}Fe_2As_2$ The measurements are carried out with incident neutron wavelength of 1.18 Å using the IN4 spectrometer at the ILL. For better visibility the experimental phonon spectra at 50 K are shifted along the y-axis by 0.04 meV-1. The calculated spectra have been convoluted with a Gaussian of FWHM of 3 meV in order to describe the effect of energy resolution in the experiment.

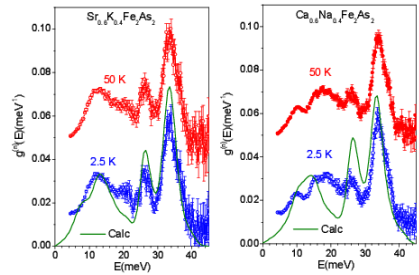

FIG. 3. (Color online) Calculated partial density of states. Solid and dashed lines correspond to $Sr_{0.6}K_{0.4}Fe_2As_2$ and $Ca_{0.6}Na_{0.4}Fe_2As_2$, respectively.

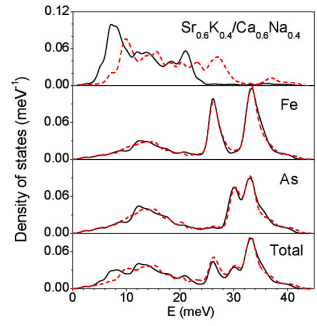

FIG. 4. (Color online) (a) The experimental phonon spectra of $Sr_{0.6}K_{0.4}Fe_2As_2$, $Ca_{0.6}Na_{0.4}Fe_2As_2$ and $BaFe_2As_2$ measured with incident neutron wavelength of 5.12 Å using IN6 spectrometer at ILL. The experimental phonon data for $BaFe_2As_2$ are taken from Ref. [12]. (b) Zoom of the low-energy part.

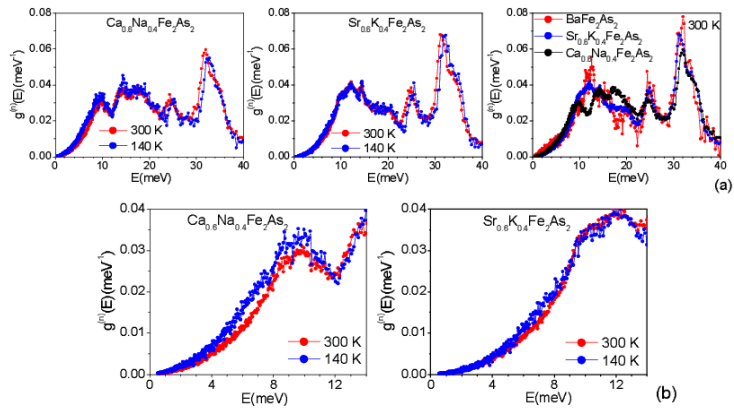